\documentclass[a4paper,11pt,preprint]{article}
\pdfoutput=1 

\usepackage{jheppub} 

\usepackage[T1]{fontenc} 
\usepackage{graphicx}

\title{\boldmath The role of radiative corrections on an extradimensional SUSY model. }

\author{J.D. Garc\'ia-Aguilar}
\author{and A. P\'erez-Lorenzana}

\affiliation{Physics Department, CINVESTAV\\Av. IPN 2508, Mexico city}

\emailAdd{jdgarcia@fis.cinvestav.mx}
\emailAdd{aplorenz@fis.cinvestav.mx}

\abstract{Theories with extra-dimensional coordinates  provide interesting mechanisms to achieve the rupture of symmetries. Here we present a novel alternative to the usual geometric considerations to achieve supersymmetric breaking for an extra-dimensional Wess-Zumino model. A supersymmetric model is constructed where the superpotential contains an effective supersymmetric non renormalizable operator, which generates, after compactification, the explicitly rupture of supersymmetry for  the excited Kaluza-Klein excitations. The supersymmetry breaking is, in turn, communicated, by the radiatives corrections, to the zero mode.}

\begin{document} 
\maketitle
\flushbottom

\section{Introduction}
\label{sec:intro}
 After the announcement by CMS and Atlas collaborations about the discovery of the Higgs boson, and from the non evidence (yet) on physics Beyond Standard Model, emerged the question of whether Supersymetry (Susy) has any real relevance on particle physics. However, in principle as there is no smoking gun to reject Susy \cite{Craig:2013cxa}, it seems that either one can reconcile the great number of MSSM parameters in order to accommodate the experiment results (PMSSM) \cite{CahillRowley:2012kx,AbdusSalam:2012ir} or make assumptions over scalars as in Split Susy models \cite{Giudice:2005wy,ArkaniHamed:2004yi}, to make it more plausible.

Supersymmetry has been motivated from various viewpoints including the gauge hierarchy problem, gauge coupling unification, and the  radiative electroweak symmetry breaking mechanism with a light Higgs boson, as well as from string theory \cite{Martin:1997ns, Murayama:2000dw}. Combining SUSY with higher-dimensional theories has also attracted a lot of attention from not only string theorists but also from phenomenologists. 

On this ground, the origin of the SUSY breaking is one of the key questions of particle physics, whose answer is yet largely unknown \cite{Luty:2005sn,Intriligator:2007cp}. In four dimensions, for example, the symmetry could be spontaneously broken in the so called hidden sector   and communicated to the visible sector by messenger fields which are singlets under SM gauge groups \cite{Giudice:1998ic}. In higher dimensional theories \cite{PerezLorenzana:2000hf,Sundrum:2005jf}, a very different geometrical view is possible: symmetries can be broken by boundary conditions on a compact space \cite{Quiros:2003gg}.

The breaking of supersymmetry by boundary conditions in compact extra dimensions has been largely explored, some examples are found in the works \cite{Mirabelli:1997aj,Kaplan:1999ac,Chacko:1999hg,ArkaniHamed:1999pv}. Some of these models provide soft breaking terms which arise through the Scherk-Schwarz mechanism \cite{Scherk:1979zr}, (for phenomenological studies see the works \cite{Barbieri:2001yz,Murayama:2012jh}). Nevertheless, the origin of this mechanism is yet unclear. The purpose of this paper is to show how this terms  could  also be generated through radiative corrections by the interactions between the zero mode and the Kaluza-Klein (KK) excited modes.

\begin{figure}[tbp]
\begin{center}
\setlength{\unitlength}{.5cm}
\begin{picture}(20,8)
\put(1,0.5){\framebox(4,6)}
\put(2.5,4){\textbf{KK}}
\put(1.9,3){\textbf{modes}}
\put(0.4,7){\emph{``Hidden sector''}}
\put(16,0.5){\framebox(4,6)}
\put(17.2,4){\textbf{Zero}}
\put(16.9,3){\textbf{modes}}
\put(15.5,7){\emph{``Visible sector''}}
\put(5,3.75){\line(1,0){3}}
\put(5,3.5){\line(1,0){3}}
\put(8.0,2.25){\framebox(5,2.5)}
\put(8.75,3.75){\textbf{Effective}}
\put(8.8,3){\textbf{coupling}}
\put(8.7,5.8){\emph{Messenger}}
\put(9.5,5.2){\emph{sector}}
\put(13,3.75){\line(1,0){3}}
\put(13,3.5){\line(1,0){3}}
\end{picture}
\caption{\label{scheme}The gap between the scalar masses and the fermion masses arises by the radiative corrections whose source are effective operators.}
\end{center}
\end{figure}
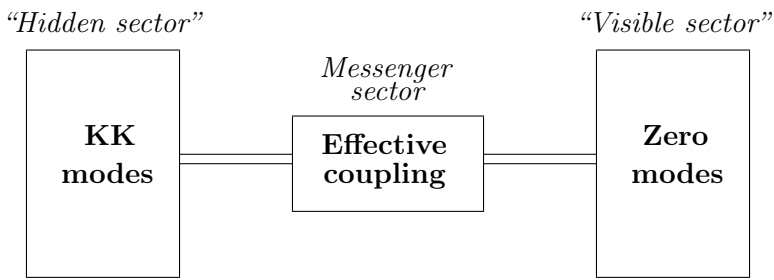

In order to make the mechanism more explicit, consider a model on compact spatial dimensions, with a compactification scale $R^{-1}$  beyond 1 TeV with non local supersymmetry. In order to achieve a chiral theory it is imposed a symmetry over the extra dimension to reduce the number of fields because the 5D model is vectorlike. Furthermore, the content of the model is restricted to include  two chiral superfields which propagates on the bulk. Thus, the inclusion of an effective operator in the superpotential leads to the rupture of supersymmetry for the excited KK modes after compactification on the orbifold $S^1/Z_2$. Additionally, the existence of interactions between zero modes and KK modes induces a rupture by radiative corrections to the zero mode supersymmetry. Remarkably, the scale is finite
 \begin{equation}
m_{SUSY}\propto \left(\frac{1}{R}\right)^2,
\end{equation}
and it is independent of the cutt-off of the theory. Schematically, the breaking mechanism which we are proposing mimics the standard idea of having a visible sector described by a SUSY invariant theory, and a hidden sector, where supersymmetry is explicitly broken, such that the breaking  of SUSY is communicated to the visible sector through some messenger that couples both sectors. 
In our mechanism all such ingredients are well defined from the 5D theory, and they can be associated to the former classification  as it is depicted in the figure~\ref{scheme}. We will elaborate a model to realize this mechanism  along this work. However, it is worth to notice that in our mechanism, it would be the coupling terms  which explicitly will break supersymmetry. 

Our paper is organized as follows. In the first part, we present a simple free model constructed under the superfield formalism that serves as the basic setup where our mechanism will be realized. We also show the off-shell model. In section 3 we introduce an effective superpotential which respects the parities assigned to the fields and the $SU(2)$ global symmetry which shall become the source of SUSY breaking. Furthermore we present the compactified model  under the orbifold $S_1/Z^2$.  Here we show how the compactification leads to  the unavoidable  rupture of supersymmetry for the Kaluza-Klein excited modes, but the zero mode preserves SUSY at tree level. Next, in section 4 we show how the susy rupture for the zero mode is achieved through the radiative corrections induced by their interactions with the KK excited modes. We found that the mass gap between a field and its super partner  is proportional to the compactification scale. Finally in section 5 we give our conclusions.	
 
\section{The model}
Here, we consider an extension of 4D Wess-Zumino model with an extra dimension, $y$, compactified on the interval $S^1/Z_2$, which is obtained upon identification under $y\rightarrow -y$ of the points of a circle $S^1$ of radius $R$.

As it is well known, the massless spectrum has $N=1$ supersymmetry in four dimensions and could be consider as the  fermion and sfermions components of the MSSM. The masive spectrum however, forms towers of Kaluza-Klein excitations for all fields living in the $5D$ bulk, with masses $n/R$ for $n=1,2,\ldots$; and they fall into supermultiplets of extended $N=2$ supersymmetry.

As pointed out in reference \cite{Mirabelli:1997aj}, in the $D=5$ the smallest spinor is a 4 component Dirac Spinor.  We follow reference \cite{ArkaniHamed:2001tb} to build the supersymmetric model in the superfields context. Thus, we take as the chiral superfields  $X=\left(A,\psi, F\right)$ and $Y=\left(B,\chi, G\right)$ where $A$, $B$ are complex scalar, $\psi$, $\chi$ are Weyl spinors and $F$ and $G$ the corresponding auxiliary fields.

Next, we notice that considering the $SU(2)$ global symmetry, it is possible to use the doublet $\Phi=\left(X~Y\right)^T$ to write the free Lagrangian as
\begin{equation}
\mathcal{L}=\left.\Phi^\dagger \Phi\right|_D+\frac{i}{2}\left. \Phi^T \sigma^2\partial_5\Phi\right|_F+h.c.,\label{mod}
\end{equation}
where $\sigma^2$ is the Pauli matrix. It is important to remark that the derivative on the fifth component appears as a superpotential, because after the compactification it is translated as the field masses.

Integrating out the auxiliary F components, the Lagrangian in (\ref{mod}) describes a $N=1$, $D=5$ supersymmetric model containing two complex scalars and one Dirac fermion 	 $\Psi=\left(\psi~\bar{\chi}\right)^T$, for which the Lagrangian reads as
\begin{equation}
\mathcal{L}=\partial_M A^\dagger \partial^M A+\partial_M B^\dagger \partial^M B+i\overline{\Psi}\Gamma^M\partial_M\Psi
\end{equation}

In order to reduce the model from five to four dimensions we use the orbifold $S^1/Z_2$.  This allows to impose the parity over superfields as 
\begin{equation}
 \left(\begin{array}{c}
 X\\Y
 \end{array}\right)\left(x^\mu,-y\right)=\left(\begin{array}{c}
 X\\-Y
 \end{array}\right)\left(x^\mu,y\right).\label{bounds}
 \end{equation}
Thereby, after  compactification, there is a $N=1$ four dimensional  massless supersymmetric  model corresponding to the zero mode of $A$ and $\psi$ fields (which make up  the chiral X superfield). This is so, due to our parity assignations. However, as it is easy to see, KK tower does preserve the global $SU(2)$ symmetry and as well as the extended $N=2$ supersymmetry. 

\section{Effective Superpotential}
The parities assigned to the superfields and the SU(2) symmetry impose constraints for  the possible operators in the superpotential. For our discussion  we consider the fourth order non renormalizable interaction term,
\begin{equation}
W\left(\Phi\right)=\frac{g}{\Lambda^2}\left(\Phi^T\Phi\right)^2+h.c.,
\end{equation}
where $\Lambda$ is the cutoff for the model. It is worth noting that such an operator respects supersymmetry. The SUSY transformations are presented on the reference \cite{Mirabelli:1997aj}, and they go as 
\begin{eqnarray}
\delta_\xi A&=&-\sqrt{2}\ \overline{\xi}^2 \Psi,\\
\delta_\xi B&=&\sqrt{2}\ \overline{\xi}^1\Psi,\\
\delta_\xi\Psi&=&i\sqrt{2}\Gamma^M\left(\xi^2\partial_M A-\xi^1\partial_M B\right)+\sqrt{2}\left(F\xi^1+G\xi^2\right),\\
\delta_\xi F&=&-i\sqrt{2}\overline{\xi}^1\Gamma\partial_M\Psi,\\
\delta_\xi G&=&-i\sqrt{2}\overline{\xi}^2\Gamma\partial_M\Psi,
\end{eqnarray}
where $\xi^1$ and $\xi^2$ are symplectic-Majorana spinors.

As usual, the interactions between fields are read off by  projection on the F-components and eliminating the auxiliary F fields. The relevant terms which interact with the zero mode are then
\begin{eqnarray}
\mathcal{L}_{int}&=&-\frac{2g}{\Lambda^2}\left(\psi\psi\left(3A^2+B^2\right)+\chi\chi A^2+4\psi\chi AB\right)\nonumber\\
&&-\frac{16g^2}{\Lambda^4}\left(\left|A\right|^6+\left|A\right|^2\left|B\right|^4+\left|A\right|^4\left|B\right|^2\right)+h.c.
\end{eqnarray}

The compactification process following the boundary conditions implied by (\ref{bounds}) leads to have an effective Wess-Zumino model for the zero mode field. Such model can be written in superfield formalism considering the chiral superfield
\begin{equation}
X_0=A_0+\sqrt{2}\theta\psi_0+\theta^2 F_0,
\end{equation}
in terms of which the zero model level Lagrangian becomes
\begin{equation}
\mathcal{L}_0=\left.X_0^\dagger X_0\right|_D+\frac{g'}{\Lambda^2}\left. X_0^4\right|_F+h.c.,
\end{equation}
where $g'=g/\left(\pi R\right)$. Therefore as already stated SUSY is preserved by the zero mode field.

On the other hand, the compactification process also implies interactions between zero mode fields and KK modes. In order to attemp writing these interaction on superfield formalism we follow the previous prescription to enconde all other fields on supermultiplets, that is
\begin{eqnarray}
X_n&=&A_n+\sqrt{2}\theta\psi_n+\theta^2 F_n;\\
Y_n&=&B_n+\sqrt{2}\theta\chi_n+\theta^2 G_n\nonumber
\end{eqnarray}
where $n=1,2,\ldots$.

As we will next show, the compactified interaction terms that involve zero and excited modes cannot always be put down in superfield formalism, evidencing the further origin of our breaking mechanism for SUSY. For example, we can take the term $-6\frac{g}{\Lambda^2}A^2\psi\psi$ which at the level of the effective KK expansion (see Appendix \ref{compacto} for an example) goes as
\begin{equation}
-\frac{6g'}{\Lambda^2}\left(A_0^2\psi_n\psi_n+A_n^2\psi_0\psi_0+4\psi_n\psi_0 A_0 A_n\right)+\ldots,
\end{equation}
notice that the first three terms on above expression can be rewritten in superfield formalism as
\begin{equation}
\frac{6g'}{\Lambda^2}\left.X_0^2 X_n^2\right|_F \label{inte}.
\end{equation}
However, by looking at purely  scalar sector, specifically the operator $-\frac{16 g^2}{\Lambda^4}\left|A\right|^6$ one can easily see that it contains the zero to KK mode couplings
\begin{equation}
 -\frac{216g'^2}{\Lambda^4} \left|A_0\right|^2\left|A_n\right|^4-\frac{144g'^2}{\Lambda^4} \left|A_0\right|^4\left|A_n\right|^2.
\end{equation}
Thus, we find that the superpotential (\ref{inte}) does not provide the correct operators to write all possible terms in the superfield formalism.

 Summarizing, we would have the effective model Lagrangian given as
 \begin{equation}
 \mathcal{L}\left(A_0,A_n,\psi_0,\psi_n\right)=\left.X_0^\dagger X_0\right|_D+\left.X_n^\dagger X_n\right|_D+\frac{6g'}{\Lambda^2}\left.X_0^2 X_n^2\right|_F-\frac{72g'^2}{\Lambda^4}\left|A_0\right|^2\left|A_n\right|^4+h.c. \label{susysoft}
 \end{equation}
 which we can see as one SUSY model plus one SUSY breaking term. There are, of course, additional non-SUSY terms involving only KK modes that we are not writing since they are not relevant for our discussion.
 
Similarly other source for a new breaking term would be the interaction between $A_0$ and $B_n$ fields.
 
 \section{Breaking zero mode supersymmetry.}
 \begin{figure}[tbp]\centering
\includegraphics[scale=0.6]{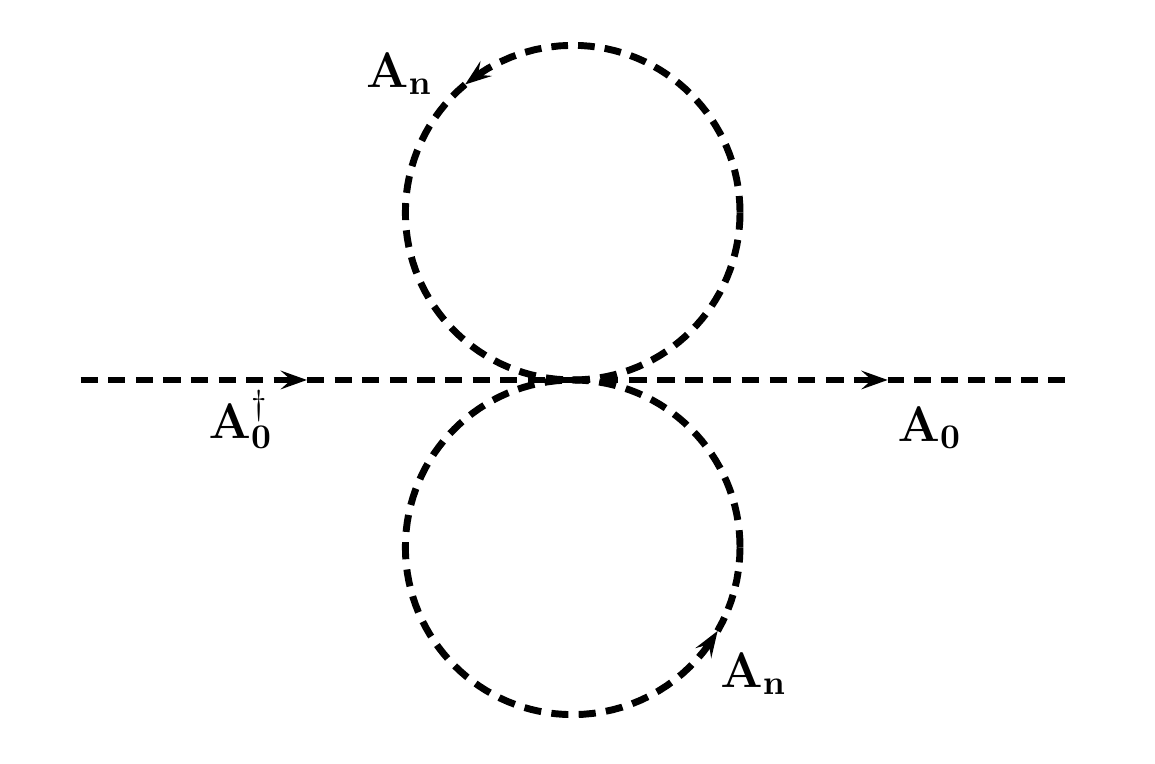}
\caption{\label{Eq:loopi}Two-loop radiative scalar mass.} 
\end{figure}
As we mentioned above the zero mode respects a global $N=1$ supersymmetry. However, the   interaction between zero mode and KK excited modes violates supersymmetry. Hence, radiative corrections will generate the rupture of the supersymmetric invariance for the zero level theory. As a consequence, a correction for the zero mode scalar field mass can be generated at the two-loop level through the diagram shown in figure~\ref{Eq:loopi},  whereas that mass for the zero mode fermion field would be protected by supersymmetry of picking up similar corrections. Thus generating an effective soft breaking term.

The contribution to the zero mode scalar mass is given by the integral
\begin{equation}
I_n\left(m\right)=-\frac{36i g'^2}{\Lambda^4}\int\frac{d^4k}{\left(2\pi\right)^4}\frac{d^4p}{\left(2\pi\right)^4}\frac{i}{k^2-m^2}\frac{i}{p^2-m^2}
\end{equation}				
 where $m^2=n^2/R^2$, which solving, can be shown to be written as 
 \begin{equation}
 I_n\left(m\right)=-\frac{9ig'^2}{4\left(2\pi\Lambda\right)^4}\left[\Lambda^2-m^2\ln\frac{\Lambda^2+m^2}{m^2}\right]^2,
 \end{equation}
 which depends on the KK index $n$ and the cutoff.
 
 At this level, it  is useful to do the  parametrization $m=x\Lambda$, where $x\in\left(0,1\right]$ to write
 \begin{equation}
 I_n\left(x\right)=-\frac{9ig'^2}{64\pi^4}\left[1-x^2\ln\frac{1+x^2}{x^2}\right]^2. \label{inti}
\end{equation} 
Notice that here the correction depends on $R$ through the effective coupling $g'$.

The result (\ref{inti}) represents the contribution due to only the interaction of one of the KK-modes. In order to obtain the total correction for the zero mode mass it is necessary to consider all KK modes whose masses are under the cut-off value. To compute that we consider the integral  
\begin{equation}
\int_0^1 \left[1-x^2\ln\frac{1+x^2}{x^2}\right]^2 dx\approx\frac{11}{25}.
\end{equation}

Last result allow us to write the total correction  for the zero mode scalar mass as
\begin{equation}
\delta m_0^2=\frac{99}{1600\pi^4}\left(\frac{g}{R}\right)^2.
\end{equation}
That means, the mass gap between scalar and fermion zero mode field shall only depend on the compactification radius, analogously to similar calculations in references \cite{Barbieri:2001yz,Murayama:2012jh}, though, associated to different models.

\section{Summary}
If Supersymmetry is present in nature, it must be broken. A theoretical mechanism to achieve this rupture should give a specific pattern   which must be in agreement with the last results by LHC and other experiments.

In this paper, we present a supersymmetric toy model on extra dimensions considering an effective superpotential that contains a generic non-renormalizable operator, which is used to generate a mass gap between field and its superpartner which is achieved, after compactification process, through radiative corrections. The so generated mass gap comes out to be inversaly proportional to compactification radius. The model as such does not require any specific scale for the compactification scale. It could be large and thus, this could explain why the last LHC results do not seem to find supersymmetry and higher dimensions.  
\appendix
\section{Compactificacion example}\label{compacto}
As we had mentioned, to achieve a 4d effective model from 5D model is necessary to apply a compactification over the extra dimension, the most common mechanism is to take a orbifold and integrate over the extra dimension. In this appendix we show an example in which we  take an real scalar  field to build the operator $g\phi^3$, and considering the Orbifold $S^1/Z_2$, also imposing the parity $Z=+1$ for $\phi$.

The effective model is given by
\begin{equation}
\mathcal{L}_{int}=g\int_0^{\pi R} \phi^3\left(x^\mu,y\right)dy.
\end{equation}
The integral is performed to remove the dependence of the Lagrangian over the extra dimension. On the other hand, parity involves to take the Kaluza-Klein decomposition as
\begin{equation}
\phi\left(x^\mu,y\right)=\frac{1}{\sqrt{\pi R}}\phi_0\left(x^\mu\right)+\sqrt{\frac{2}{\pi R}}\sum_{n=1}\phi_n\left(x^\mu\right)\cos\left(\frac{ny}{R}\right)
\end{equation}
where $\phi_n\left(x^\mu\right)$ is called the KK mode.

These considerations lead to consider the integral
\begin{equation}
\int_0^{\pi R} \cos \left(\frac{ny}{R}\right) \cos \left(\frac{py}{R}\right)\cos \left(\frac{qy}{R}\right) dy=\frac{\pi R}{4}\left(\delta_n^{~-\left(q+p\right)}+\delta_n^{~q-p}+\delta_n^{~p-q}+\delta_n^{q+p}\right)
\end{equation}
where $\delta_x^y$ is the Kronecker delta and represents the KK number conservation. 

The KK number conservation restrict the effective operators, for example the zero mode ($\phi_0$) interacts with the KK modes with the Lagrangian given by
\begin{equation}
\mathcal{L'}_{int}=g'\phi_0\left(\phi_0^2+\phi_n^2\right)
\end{equation}
 where $g'=g/\sqrt{\pi R}$.

\acknowledgments

This work was supported in part by Conacyt, M\'exico, under grant No. 132061.

\end{document}